# Topic Map: An Ontology Framework for Information Retrieval

Rajkumar Kannan

Department of Computer Science
Bishop Heber College(Autonomous)
Tiruchirappalli, TN, India. www.bhc.edu.in
*rajkumar@bhc.edu.in*

*Abstract -* **The basic classification techniques for organizing information are thesauri, taxonomy and faceted classification. Topic map is relatively a new entrant to this information space. Topic map standard describes how complex relationships between abstract concepts and real world resources can be represented using XML syntax. This paper explores how topic map incorporates the traditional techniques and what are its advantages and disadvantages in several dimensions such as content management, indexing, knowledge representation, constraint specification and query languages in the context of information retrieval. The constructs of topic maps are illustrated with a use-case implemented in XTM.**

## I.   INTRODUCTION

Unable to find the information which exists in a document object or collection of objects is a traditional problem in information retrieval. There are age-old traditional techniques that find the result by classifying objects through its subjects. The important subject-based classification methods [1] are as follows:

*Controlled vocabulary* is a set of indexing terms or subjects used for classification. *Taxonomy* is a subject-based classification technique that arranges the terms in the controlled vocabulary into a hierarchy. Thereby, related terms can be grouped together and categorized for efficient search. But, search is very limited here as it cannot capture several relationships and synonyms, which are required to reduce the search space.

*Thesauri* basically extend taxonomy to describe the world by not only allowing subjects to be arranged in a hierarchy, but also allowing other statements to be made about the subjects. It includes properties such as broader term, scope note, synonym term, root (top most) term and related term. In short, thesauri provide a much richer vocabulary for describing the terms than taxonomies do, and so are much more powerful tools. As can be seen, using a thesaurus instead of taxonomy would solve several practical problems in classifying objects and also in searching for them.

*Facets* can be thought of as different axes along which documents can be classified, and each facet contains a number of terms. In faceted classification the idea is to classify documents by picking one term from each facet to describe the document along all the different axes. This would then describe the document from many different perspectives.

*Ontology* is a model for describing the world that consists of a set of types, properties, and relationship types. Controlled vocabulary, taxonomy, thesauri, and facets are fixed vocabulary languages where as ontology is a open vocabulary language.

Topic maps emerged as a way of merging of electronic indexes. A topic map consists of a collection of *topics*, each representing some real world thing or concept, as for example terms in an index of a book refer concepts inside a book through page numbers defined in the book index. Topics are related to each other by *associations*, which are typed n-ary combinations of topics. A topic may also be related to any number of resources by its *occurrences*. The important benefits [2] are

- Separation of topic space from resource space
- No fixed ontology
- XML syntax
- Flexible schema

- Flexible extension

In this paper, we will introduce the fundamental constructs of topic map which are topic, association and occurrences and show how the concept of scope can give additional power to the model. In order to explain the power of subject-based classification through topic maps, we will consider *university faculty profile* as a subject of study and illustrate the various topics and their associations in XTM syntax.

Rest of the paper is organized as follows. Section2 introduces TAO of topic maps. Topic names and its types are explained in section 3 and 4 respectively. The occurrence of information resources are discussed in section 5, while section 6 explains how as association relates topics. Finally, section 7 concludes the paper.

## II. TAO OF TOPIC MAP

Topic map is a network of nodes, not a tree hierarchy, representing a collection of topics, associations, occurrences and scope. (see. Figure1).

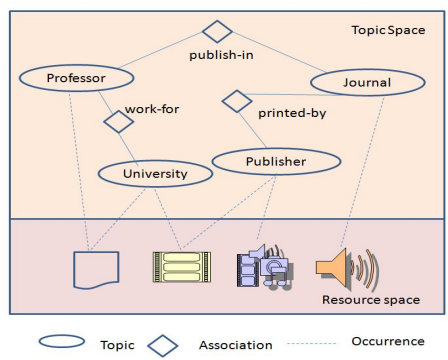

Figure 1. TAO of Topic Maps.

Figure 1 shows the three fundamentals of topic map: topics, associations and occurrences. Here, topic map universe is divided into topic space and resource space through topic-to-topic and topic-to-resource relationships. This partitioning helps merging and extending topic map alone without affecting information resources. Also, information resources can be updated without disturbing topic map.

## III. THE NAMES OF TOPICS

In topic maps, a topic can be given one or more names so all are synonyms. This feature facilitates us to query based on not only its topic name but also on its synonym list. So entire topic map is populated with synonym lists and now the topic map becomes synonym ontology enabling knowledge based querying on topic maps. Figure 2 depicts

XTM syntax [3] for topics, *Professor* and *rajkumar-kannan*.

```
<topic id="#professor">
<baseName>
<baseNameString>Professor</ba
seNameString>
</baseName>
</topic>
<topic id="#rajkumar-kannan">
<instanceOf>
<topicRef
xlink:href="#professor"/>
</instanceOf>
<baseName>
<baseNameString>Rajkumar
Kannan</baseNameString>
</baseName>
</topic>
```

Figure 2. Topics Professor and Rajkumar-Kannan

Further, different topics in a topic map can have same name (i.e., duplicates), which taxonomies and thesauri do not allow. To resolve this, topic maps use its type, occurrences and associations. For example, home page of a web site can be identified by its URI and also through the menu, such as *About* menu. For example, the topic name, Dr Rajkumar Kannan is constrained to the topic, university only as shown in figure3.

```
<baseName>
<scope>
<topicRef
xlink:href="#university"/>
</scope>
<baseNameString>Dr  Rajkumar
Kannan</baseNameString>
</baseName>
```

Figure 3. Scope of Dr-Rajkumar-Kannan

Topic name can be given a *scope*, which is a set of topics representing the context (or constraint) in which the name is appropriate [4, 5]. This enables IR system to delimit the search space. For instance, Company Region can be scoped with Europe, Asia and so on. Therefore, users by selecting a region, IR system will display the results according. Further, scope of a topic in a topic may is quite popular to support multilingual retrieval. Since scope consists of topics, topic map can be navigated even through scope for better searching.

## IV. TYPES OF TOPICS

In topic maps topics can be given its type, an important feature that is missing from traditional classification techniques. Suppose we run a query *Tirupathi* on traditional classification based system, it will return all results about the city in India by name *Tirupathi* and all persons whose names are *Tirupathi*. If we are interested only about the city, not the persons, we have no way to describe in these classification systems. Through topic types we can specify *"Tirupathi" is a "city"* where city is a topic type of a topic *Tirupathi*. Thereby users can perform a search such as "find 'thirupathi', but show only 'places'". Also, remember since the types of topics are themselves topics the designer of the topic map can choose which types to use [6,]

## V. OCCURRENCES OF INFORMATION RESOURCES

Occurrences connect topics to information resources that contain information about them. For example, page number in a book index indicates where the information about the subject could be found inside book. Occurrences are resources that are either addressable by reference using a URI or capable of being placed inline as string. In figure4, a journal topic is associated with the information resource, NCAKM10-paper by occurrence tag.

```
<topic id="#NCAKM10-paper">
<instanceOf><topicRef
xlink:href="#journal"/></instanceOf>
<baseName>
<baseNameString>Advances in
Knowledge Management special Issue
</baseNameString>
</baseName>
<occurrence>
<instanceOf><topicRef xlink:href="#pdf-
format"/></instanceOf>
<resourceRef    xlink:href=    http://
www.rajkumarkannan.org    /    pub
/ncakm10.pdf/>
</occurrence>
</topic>
```

Figure 4.    Information resource defined using occurrence tag.

Since occurrences can have type, it indicates what the resource is. For instance, occurrences indicate whether it is conference paper, journal paper, video clip of a keynote, or slides of an oral presentation in a conference. So we can differentiate occurrences so that the query can isolate the search space. Occurrences can have scope too. Therefore a scope, *refereed conference paper*, distinguishes conference papers which are from reputed conferences from non-refereed. Since occurrence types are topics, and so the designer is free to define occurrence types at his will. This open vocabulary feature does not exist in traditional techniques.

## VI. ASSOCIATION BETWEEN TOPICS

An association is a relationship between topics. Each association plays a role as a member of that association. The roles a topic plays in associations can be typed and scoped [7, 8]. There is no directionality inherent in an association. For instance Figure5 illustrates the association "works-for" for the professor, rajkumar-kannan. Note that the role played by him is teaching. Traditional classification schemes provide only very little features such as broader-narrower terms. However, topic maps precisely define the associations and hence users' query will correctly retrieve the required result.

```
<association>
<instanceOf>
<topicRef xlink:href="#works-
for"/>
</instanceOf>
<member>
<roleSpec>
<topicRef
xlink:href="#teaching/>
</roleSpec>
<topicRef
xlink:href="#rajkumar-
kannan"/>
</member>
</association>
```

Figure 5.    Works-for association.

Finally, topic map is a collection of topics. Therefore the topic map for Figure1 will include all topics, associations and occurrences. Figure6 shows a segment of topic map for professor topic map

```
<topicMapxmlns=http://www.top
icmaps.org/xtm/1.0/
xmlns:xlink="http://www.w3.org
/1999/xlink">
        <!--all        topics,
associations defined here.. --
>
</topicMap>
```

Figure 6.    Segment of professor topic map for Figure1.

## VII. CONCLUSION

In this article, we have introduced topic maps technology as an ISO standard. We presented the fundamental building blocks of topic map model, showing its power to efficiently search for the required objects through a use-case of faculty profile defined in XTM syntax.

## REFERENCES

[1] Lars Marius Garshol, Metadata, Thesauri, Taxonomies and Topic Maps-Making sense of it all, Journal of Information Science, 30(4):378-391, 2004; ISSN 0165-5515.

[2] Kal Ahmed and Graham Moore, An Introduction to Topic Maps, IBM Systems Journal, July 2005

[3] Moore G., Pepper S. (ed.), XML Topic Maps (XTM) 1.0 [online], TopicMaps.Org. HTML format.

[4] http://www.topicmap.org

[5] http://www.ontopia.org

[6] Ahmed K., TMShare—Topic Map Fragment Exchange in a Peer-To-Peer Application. HTML format.

[7] Ahmed K., Topic Map Design Patterns for Information Architecture. HTML format.

[8] http://www.isotopicmaps.org/